\documentclass[a4paper,12pt]{article}
\usepackage{amsmath}
\usepackage{bm}
\usepackage{authblk}

\numberwithin{equation}{section}

\DeclareMathOperator{\Realpart}{Re}

\DeclareMathOperator{\sgn}{sgn}

\title{Directional dependence of the Unruh effect\\ 
for spatially extended detectors}

\author{Sanved Kolekar\footnote{sanved.kolekar@cbs.ac.in} \,}

\affil{ UM-DAE Centre for Excellence in Basic Sciences, \\  Mumbai 400098, India}

\date{November 2019}
\begin{document}
\maketitle

\begin{abstract}
We analyse the response of a spatially extended direction-dependent local quantum system, a detector, 
moving on the Rindler trajectory of uniform linear acceleration in Minkowski spacetime, and coupled linearly to a quantum scalar field. 
We consider two spatial profiles: (i) a profile defined in the Fermi-Walker frame of an arbitrarily-accelerating trajectory, generalising the isotropic Lorentz-function profile introduced by Schlicht to include directional dependence; and 
(ii) a profile defined only for a Rindler trajectory, utilising the associated frame, and confined to a Rindler wedge, but again allowing arbitrary directional dependence. For (i), we find that the transition rate on a Rindler trajectory is non-thermal, and dependent on the direction, but thermality is restored in the low and high frequency regimes, with a direction-dependent temperature, and also in the regime of high acceleration compared with the detector's inverse size. For~(ii), the transition rate is isotropic, and thermal in the usual Unruh temperature.
We attribute the non-thermality and anisotropy found in (i) to the \textit{leaking} 
of the Lorentz-function profile past the Rindler horizon. 
\end{abstract}

\section{Introduction}

The Unruh effect \cite{Fulling:1972md,Davies:1974th,Unruh:1976db}
states that an observer of negligible spatial size on a worldline 
of uniform linear acceleration in Minkowski spacetime 
reacts to the Minkowski vacuum of a relativistic quantum field 
by thermal excitations and de-excitations, in the Unruh temperature~$g/(2\pi)$, 
where $g$ is the observer's proper acceleration. The acceleration singles out a distinguished direction in space, 
and an observer with direction-sensitive equipment will in general see a direction-dependent response; 
however, for the Lorentz-invariant notion of direction-sensitivity introduced in~\cite{Takagi:1985tf}, 
the associated temperature still turns out to be equal to~$g/(2\pi)$, 
independently of the direction. 
For textbooks and reviews, see~\cite{Birrell:1982ix,Crispino:2007eb,Fulling:2014wzx}. 

In this paper we address the response of uniformly linearly accelerated observers in Minkowski spacetime, 
operating direction-sensitive equipment of nonzero spatial size. 
We ask whether the temperature seen by these observers is 
still independent of the direction. 
The question is nontrivial: while 
a spatially pointlike detector with a monopole coupling is known to be a good approximation 
for the interaction between the quantum electromagnetic field and electrons 
on atomic orbitals in processes where the angular momentum interchange 
is insignificant \cite{MartinMartinez:2012th,Alhambra:2013uja}, 
finite size effects can be expected to 
have a significant role in more general situations 
\cite{DeBievre:2006pys,Hummer:2015xaa,Pozas-Kerstjens:2015gta,Pozas-Kerstjens:2016rsh,Pozas-Kerstjens:2017xjr,Simidzija:2018ddw}. 
Also, the notion of a finite size accelerating body has significant subtlety: while a rigid body undergoing 
uniform linear acceleration in Minkowski spacetime can be defined in terms of the boost Killing vector, different points on the body have differing values of the proper acceleration, and the body as a whole does not have an unambiguous value of `acceleration'. It would be interesting to ask whether the resultant transition rate would be thermal at all. And if yes, then at what temperature. The observer's response could hence well be expected to depend explicitly on the body's shape as well as the size. 

A related issue is the following: An analysis of a direction dependent point-like detector re-affirms that the Unruh bath is isotropic even though there is a preferred direction in the Rindler frame, the spatial direction along the direction of acceleration~\cite{Takagi:1985tf}. However, analysing drag forces on drifting particles in the Unruh bath reveals through the Fluctuation Dissipation Theorem that the quantum fluctuations in the Unruh bath are not isotropic~\cite{Kolekar:2012sf}. These anisotropies could be relevant for direction dependent spatially extended detectors whose length scales are of the order or greater than the correlation scales associated with the quantum fluctuations.

We consider the response of spatially extended direction-dependent detectors in uniform linear acceleration in two models of such a detector: (i) a spatial sensitivity profile that generalises the isotropic Lorentz-function considered by Schlicht \cite{schlicht} to include spatial anisotropy, and (ii) a spatial sensitivity profile defined in terms of the geometry of the Rindler wedge, and explicitly confined to this wedge, following De Bievre and Merkli~\cite{DeBievre:2006pys}. 

We begin in Section \ref{schlichtsection} by briefly reviewing a detector with an isotropic Lorentz-function spatial profile~\cite{schlicht}, highlighting the role of a spatial profile as the regulator of the quantum field's Wightman function, and recalling how the Unruh effect thermality arises for this detector. 

In section \ref{directiondetsection} we generalise the Lorentz-function spatial profile to include spatial anisotropy, 
initially for an arbitrarily-accelerated worldline, relying on the Fermi-Walker frame along the trajectory. We then specialise to a Rindler trajectory of uniform linear acceleration. We find that the transition rate is non-thermal, and angle dependent.
Thermality is however restored in the low and high frequency regimes, and also in the regime of high acceleration compared with the inverse of the detector's spatial extent. 

In section \ref{rindlerframesection} we analyse a profile defined in the Rindler frame of a Rindler trajectory, and confined to the Rindler wedge, following De Bievre and Merkli~\cite{DeBievre:2006pys}. We find that the transition rate is isotropic and thermal at the usual Unruh temperature. 

In section~\ref{discsection} we discuss and resolve the discrepancy of these two outcomes. The key property responsible for the non-thermality and anisotropy for the Lorentz-function profile is that this profile leaks outside the Rindler wedge, past the Rindler horizon. 
The leaking is an unphysical side effect of a detector model with a noncompact spatial profile, 
and it is unlikely to have a counterpart in spatially extended detectors with a more fundamental microphysical description. We leave the development of such spatially extended detector models subject to future work.

\section{Spatially isotropic Lorentz-function profile\label{schlichtsection}}

In this section we briefly review Schlicht's generalisation \cite{schlicht} 
of a two-level Unruh-DeWitt detector \cite{Unruh:1976db,DeWitt:1979} to a nonzero spatial size. 

We consider a massless scalar field $\phi$ in four-dimensional Minkowski spacetime, 
and a two-level quantum system, a detector, localised around a timelike worldline $x(\tau)$, parametrised in terms of the proper time~$\tau$. 
The interaction Hamiltonian reads 
$H_{int} = c \, m(\tau) \,\chi(\tau) \phi(\tau)$, where $c$ is a coupling constant, $m(\tau)$ is the detector's monopole moment operator, 
$\chi(\tau)$ is the switching function that specifies how the interaction is turned on and off, and $\phi(\tau)$ is the spatially smeared field operator. The formula for $\phi(\tau)$ is 
\begin{equation}
\phi(\tau) = \int d^3\xi \; f_{\epsilon} ({\bm \xi}) \, \phi(x(\tau,  {\bm \xi}))
\ , 
\label{smearedoperator}
\end{equation}
where ${\bm \xi} = (\xi^1, \xi^2, \xi^3)$ stands for the spatial coordinates associated with the local 
Fermi-Walker transported frame and $x(\tau, {\bm \xi})$ is a spacetime point written in terms of the Fermi-Walker coordinates. 
The smearing profile function $f_{\epsilon} ({\bm \xi})$ specifies the spatial size and shape of the detector in its instantaneous rest frame. 
In linear order perturbation theory, the detector's transition probability is then proportional to the response function, 
\begin{equation}
{\cal F}(\omega) = \int_{-\infty}^{\infty} du \, \chi(u)  \int_{-\infty}^{\infty} ds \, \chi(u -s) e^{- i \omega s} \, W(u,u-s)
\ , 
\label{transprobability}
\end{equation}
where $\omega$ is the transition energy, 
$W(\tau,\tau^\prime) = \langle \Psi | \phi(\tau) \phi(\tau^\prime) | \Psi \rangle$ and $|\Psi \rangle$ is the initial state of the scalar field. 
The choice for the smearing profile function $f_{\epsilon}$ made in \cite{schlicht} was the three-dimensional isotropic Lorentz-function, 
\begin{equation}
f_{\epsilon}({\bm \xi})= \frac{1}{\pi^2} \frac{\epsilon}{{(\xi^{2}+\epsilon^{2})}^2} 
\ , 
\label{schlichtprofile}
\end{equation} 
where the positive parameter $\epsilon$ of dimension length characterises the effective size. 

The selling point of the profile function is that it allows the switch-on and switch-off to be made instantaneous; 
for a strictly pointlike detector,  by contrast, 
instantaneous switchings would produce infinities and ambiguities~\cite{schlicht}. 
In particular, for a detector that is switched off at proper time $\tau$, the derivative of ${\cal F}$ with respect to $\tau$ can be understood as a transition rate, in the `ensemble of ensembles' sense discussed in~\cite{Langlois:2005if,Louko:2007mu}. 
If the switch-on takes place in the infinite past, the transition rate formula becomes 
\begin{equation}
{\dot {\cal F}}(\omega) =  2 \Realpart \int_0^\infty ds \,e^{-i \omega s} \, W(\tau,\tau - s)
\ . 
\label{eq:transrate-schlicht}
\end{equation}

When the trajectory is the 
Rindler trajectory of uniform linear acceleration of magnitude $g>0$, 
and $|\Psi\rangle$ is the Minkowski vacuum, 
the transition rate \eqref{eq:transrate-schlicht} becomes \cite{schlicht} 
\begin{equation}
{\dot {\cal F}}(\omega) =  \frac{1}{2 \pi} \; \frac{(\omega /g)}{1+ \epsilon^2} \;  \frac { e^{\frac{2\omega }{g} \tan^{-1}\left( g \epsilon \right)}}
{
e^{\frac{2 \pi \omega}{g}} -1
}
\ . 
\label{schlichttrans}
\end{equation}
In the limit $\epsilon \rightarrow 0$, ${\dot {\cal F}}$ \eqref{schlichttrans} 
reduces to the Planckian formula in the Unruh temperature~$g/(2\pi)$, 
consistently with other ways of obtaining the response of a pointlike detector in the long time limit 
\cite{Unruh:1976db,DeWitt:1979,letaw,Takagi:1986kn,Fewster:2016ewy}. 

For $\epsilon$ strictly positive, ${\dot {\cal F}}$ \eqref{schlichttrans} is no longer Planckian. 
However, we wish to observe here that ${\dot {\cal F}}$ \eqref{schlichttrans} is still thermal, 
in the sense that it satisfies the detailed balance condition, 
\begin{align}
{\dot {\cal F}}(-\omega) = e^{\beta\omega}{\dot {\cal F}}(\omega)
\ , 
\end{align}
where the inverse temperature now reads $\beta = \bigl(2\pi - 4 \tan^{-1}( g \epsilon) \bigr) /g$. 
The temperature is thus higher than the usual Unruh temperature. 
This feature has to our knowledge not received attention in the literature, and we shall 
discuss its geometric origins in section~\ref{discsection}.

\section{Spatially anisotropic Lorentz-function profile}\label{directiondetsection}

In this section we generalise the isotropic Lorentz-function profile \eqref{schlichtprofile} to include spatial anisotropy. 

\subsection{General trajectory}

Let $x(\tau)$ again be a timelike worldline parametrised by its proper time $\tau$, 
so that the four-velocity $u^a := \frac{dx^a}{d\tau}$ is a unit timelike vector. 
The four-acceleration vector $a^a := u^b \nabla_b u^a$ is orthogonal to $u^a$, 
and its direction is Fermi-Walker transported along the trajectory only when the trajectory stays in a timelike plane, as seen by considering the torsion and hypertorsion of the trajectory in the Letaw-Frenet equations \cite{letaw, kolekar}.

We define the direction dependence by writing the expression for $\phi(\tau)$ in Eq.(\ref{smearedoperator}) as
\begin{equation}
\frac{d \phi(\tau)}{d \Omega} = \int d\xi \; f_{\epsilon} ({\bm \xi}) \, \phi(x(\tau,  {\bm \xi})) = \phi_\Omega(\tau)
\ , 
\label{smearedoperatorang}
\end{equation}
such that the ${\bm \xi}$ points in the direction of $\Theta_0$ and $\phi_0$.  Integrating $\phi_\Omega(\tau)$ all over the solid angle then reproduces the smeared field operator $\phi(\tau)$ in the Schlicht case. Assuming the two level quantum system to couple linearly to $\phi_\Omega$, one can then proceed to calculate the transition rate as per formula in Eq.(\ref{transprobability}) with the corresponding $W_\Omega(\tau,\tau^\prime)$ equal to $ \langle \Psi | \phi_\Omega(\tau) \phi_\Omega(\tau^\prime) | \Psi \rangle$. Equivalently, one can consider a detector whose spatial profile has the radial dependence of \eqref{schlichtprofile} and depends on the angles $\Theta_0$ and $\phi_0$ through
\begin{align}
f_{\epsilon}({\bm \xi},\Theta_{0})= \frac{1}{2\pi^3} \frac{\epsilon}{{(\xi^{2}+\epsilon^{2})}^2} \frac{\delta (\theta - \Theta_{0})}{\sin\Theta_{0}} \delta(\phi - \phi_0)
\ , 
\label{angprof}
\end{align}
where $\theta$ and $\phi$ are measured in the ${\bm \xi} = (\xi^1, \xi^2, \xi^3)$ space. 
One can once again note that integrating \eqref{angprof} over the solid angle $d\Omega_0 = \sin\Theta_{0} d\Theta_{0} d\phi_{0}$ yields the isotropic profile~\eqref{schlichtprofile}. 

Following the steps in section~\ref{schlichtsection}, the transition rate formula becomes 
\begin{align}
{\dot {\cal F}}_{\Theta_0}(\omega) =  2 \Realpart \int_0^\infty ds \,e^{-i \omega s} \, W_{\Theta_0}(\tau,\tau - s)
\ , 
\label{angtransitionrate}
\end{align}
where 
\begin{align}
W_{\Theta_0}(\tau,\tau^\prime) = \langle \Psi | \phi(\tau, \Theta_0) \phi(\tau^\prime,\Theta_0) | \Psi \rangle
\ , 
\label{angwhitmannfunction}
\end{align}
and the smeared field operator reads 
\begin{align}
\phi(\tau, \Theta_0) = \int d^3\xi \; f_{\epsilon} ({\bm \xi}, \Theta_0) \, \phi(x(\tau,  {\bm \xi}))
\ , 
\label{smearedRARF}
\end{align}
provided these expressions are well defined. 
We shall now show that the expressions are well defined provided $\Theta_0 \ne \pi/2$, 
and we give a more explicit formula for ${\dot {\cal F}}_{\Theta_0}$. 

Suppose hence from now on that $\Theta_0 \ne \pi/2$. 
We work in global Minkowski coordinates in which points on 
Minkowski spacetime are represented by their position vectors, 
following the notation in~\cite{schlicht}. 
The trajectory is written as~$x^b(\tau)$. 
At each point on the trajectory, we introduce three spacelike unit vectors 
$e^b_{(\alpha)}(\tau)$, $\alpha = 1,2,3$, which are orthogonal to each other and to 
$u^b(\tau) = \frac{dx^b(\tau)}{d\tau}$, and are Fermi-Walker transported along the trajectory. 
We coordinatise the 
hyperplane orthogonal to $u^b(\tau)$ by ${\bm \xi} = (\xi^1, \xi^2, \xi^3)$ by 
\begin{align}
x^b(\tau, {\bf \xi}) = x^b(\tau) + \xi^\alpha e^b_{(\alpha)}(\tau)
\ . 
\end{align}

Using (\ref{angwhitmannfunction}) and (\ref{smearedRARF}), we obtain 
\begin{equation}
W_{\Theta_0}(\tau,\tau^\prime) = \frac{1}{(2 \pi)^3} \int \frac{d^3 {\bf k}}{2 \omega({\bf k})} \;  
g_{\Theta_0} ( {\bf k}, \tau) \, g^*_{\Theta_0} ( {\bf k}, \tau^\prime)
\ , 
\label{gwhitmann}
\end{equation}
where $\omega({\bf k}) = \sqrt{{\bf k}^2}$ and 
\begin{equation}
g_{\Theta_0} ( {\bf k}, \tau) = \int d^3\xi \; f_{\epsilon}({\bm \xi},\Theta_{0}) e^{i k_b x^b(\tau, \, {\bm \xi} )}
\ .  
\label{gdef}
\end{equation}
The index $\alpha$ refers to values $(1,2,3)$ and $e^b_{(\alpha)}(\tau)$ are the orthogonal Fermi unit-basis vectors in the spatial direction orthogonal to $u^b(\tau)$. Then defining $3$ - vector $\tilde{{\bf k}}$ having components $(\tilde{{\bf k}})_\alpha = k_b {\bf e}^b_{(\alpha)}(\tau)$ and working in spherical co-ordinates in the ${\bf \xi}$- space, we can recast Eq.~(\ref{gdef}) to get
\begin{eqnarray}
g_{\Theta_0}( {\bf k}, \tau)  &=& \frac{1}{\pi} e^{i k_b x^b(\tau)} \int_0^\infty d\xi \, \frac{\xi^2\epsilon}{(\xi^2+\epsilon^2)^2} e^{i \xi \cos\Theta_0 |\tilde{{\bf k}}|} 
\notag 
\\
& = & e^{i k_b x^b(\tau)} \left( I_R + I_M \right)
\ , 
\end{eqnarray}
where $I_R$ and $I_M$ are the real and imaginary parts of the integral. Here, $\tilde{{\bf k}}$ is oriented along the $z$ direction in the ${\bf \xi}$- space and the angle $\Theta_0$ is measured from the $z$ direction. 
The real part can be evaluated by contour integration, with the result
\begin{eqnarray}
I_R &=& \frac{1}{\pi} \int_{0}^\infty d\xi \frac{\xi^2\epsilon}{(\xi^2+\epsilon^2)^2} \cos(\xi |\tilde{{\bf k}}|   \cos\Theta_0) \nonumber \\
&=& \frac{1}{4} \frac{\partial}{\partial \epsilon} \left( \epsilon e^ {- \epsilon |\tilde{{\bf k}}| |\cos{\Theta_0}| }\right) \nonumber \\
&=& \frac{1}{4} \left( 1-\epsilon |\tilde{{\bf k}}|  |\cos\Theta_0| \right) e^{-\epsilon |\tilde{{\bf k}}|  |\cos \Theta_0|}
\ . 
\label{IR}
\end{eqnarray}
The imaginary part can be reduced to the exponential integral $E_1$~\cite{dlmf}, with the result 
\begin{align}
I_M = \frac{i \sgn(a)}{4\pi} 
\Bigl[ 
(|a|+1) e^{|a|} E_1(|a|) + (|a|-1) e^{-|a|} \Realpart \bigl( E_1(-|a|) \bigr) 
\Bigr] 
\ , 
\end{align}
where $a = \epsilon |\tilde{{\bf k}}| \cos \Theta_0$. 
Note that the replacement $\Theta_0 \to \pi - \Theta_0$ leaves $I_R$ invariant but gives $I_M$ a minus sign. 

To proceed further, we assume that the profile function is invariant under $\Theta_0 \rightarrow \pi - \Theta_0$, 
that is, under $\cos(\Theta_0) \rightarrow  - \cos(\Theta_0)$. 
Since $I_M$ is an odd function of $\cos(\Theta_0)$, it does not contribute to $g_{\Theta_0} \left( {\bf k}, \tau \right)$ under such an invariance whereas $I_R$ being even in $\cos(\Theta_0)$ contributes. Physically, this would mean that the direction sensitive detector reads off the average of two transition rates from the $\Theta_0$ and $\pi - \Theta_0$ directions respectively. Eq.(\ref{gdef}) then becomes
\begin{equation}
g_{\Theta_0} \left( {\bf k}, \tau \right) = \frac{1}{4} \frac{\partial}{\partial \epsilon} \left( \epsilon \; e^ {- \epsilon |\tilde{{\bf k}}| |\cos{\Theta_0}| } \; e^{i k_b x^b(\tau)}\right)
\end{equation}
Using the fact that $k_b$ is a null vector, it is straightforward to show that $|\tilde{{\bf k}}| = - [k_b u^b(\tau)]$. Substituting the above expression in Eq.(\ref{gwhitmann}) and upon performing the straightforward ${\bf k}$ integral, we can write a compact expression for $W_{\Theta_0}$ of the following form 
\begin{eqnarray}
W_{\Theta_0}(\tau,\tau^\prime) &=& \frac{-1}{16} \frac{\partial}{\partial \epsilon^\prime} \frac{\partial}{\partial \epsilon^{\prime \prime}} \Biggl( \frac{4\pi\epsilon^\prime \epsilon^{\prime \prime} }{ \left[ T\left(\epsilon^\prime, \epsilon^{\prime \prime}, \tau, \tau^\prime  \right) \right]^2  -  \left[ X \left( \epsilon^\prime, \epsilon^{\prime \prime}, \tau, \tau^\prime  \right) \right]^2 } \Biggr)_{\epsilon^\prime = \epsilon, \epsilon^{\prime \prime} = \epsilon} 
\label{wfinalcompact}
\end{eqnarray}
where the functions $T\left(\epsilon^\prime, \epsilon^{\prime \prime}, \tau, \tau^\prime  \right)$ and $X\left(\epsilon^\prime, \epsilon^{\prime \prime}, \tau, \tau^\prime  \right)$ are found to be
\begin{eqnarray}
T\left(\epsilon^\prime, \epsilon^{\prime \prime}, \tau, \tau^\prime  \right) &=&    \left(t(\tau) -t(\tau^\prime) \right) - i |\cos\Theta_{0}| \left( \epsilon^{\prime \prime} {\dot{t}}(\tau)  + \epsilon^{\prime} {\dot{t}}(\tau^\prime) \right) \label{Tdef} \\
X\left(\epsilon^\prime, \epsilon^{\prime \prime}, \tau, \tau^\prime  \right) &=&   \left({\bf x}(\tau) -{\bf x}(\tau^\prime) \right) - i |\cos\Theta_{0}| \left( \epsilon^{\prime \prime} {\dot{{\bf x}}}(\tau)  + \epsilon^{\prime} {\dot{{\bf x}}}(\tau^\prime) \right) \label{Xdef}
\end{eqnarray}
The overdot refers to the derivative with respect to the proper time $\tau$ or $\tau^\prime$. 
Expanding the above expression,  $W_{\Theta_0}(\tau,\tau^\prime) $ can also be written as
\begin{eqnarray}
W_{\Theta_0}(\tau,\tau^\prime)  = \frac{1}{16} & \bigg\{& \frac{4 \pi}{-T^2 + X^2} +  \frac{i 8 \pi \epsilon^{\prime \prime} |\cos\Theta_{0}| \left[ -T\dot{t} + X \dot{{\bf x}} \right] }{\left( -T^2+X^2 \right)^2} \nonumber \\
&& + \frac{i 8 \pi \epsilon^{\prime} |\cos\Theta_{0}| \left[-T \dot{t}^{\prime} + X \dot{{\bf x}}^{\prime} \right]}{\left(-T^2+X^2 \right)^2} \nonumber \\
&& - \frac{32 \pi \epsilon^{\prime} \epsilon^{\prime\prime} |\cos\Theta_{0}| \left[ -T \dot{t}^\prime + X \dot{{\bf x}}^{\prime} \right] \left[ -T \dot{t}+X\dot{{\bf x}} \right] }{\left( -T^2+X^2 \right)^3} \nonumber \\ 
&& + \frac{8 \pi \epsilon^{\prime} \epsilon^{\prime\prime} |\cos\Theta_{0}|^{2} \left[- \dot{t} \dot{t}^{\prime} + \dot{{\bf x}}\dot{{\bf x}}^{\prime} \right] }{ \left( -T^2+X^2 \right)^2} \; \bigg\}_{\epsilon^\prime = \epsilon, \epsilon^{\prime \prime} = \epsilon} 
\label{Wfinal}
\end{eqnarray}
The first term is of the familiar form one gets for the total transition rate, however one must note that both the functions $T\left(\epsilon^\prime, \epsilon^{\prime \prime}, \tau, \tau^\prime  \right)$ and $X\left(\epsilon^\prime, \epsilon^{\prime \prime}, \tau, \tau^\prime  \right)$ are dependent on the angle $\Theta$. 

Another feature of Eq.(\ref{Wfinal}), is that, $\epsilon$ and $|\cos\Theta_0|$ always appear as a product in the expression. Given that $|\cos\Theta_0|$ is always non-negative, one can formally absorb it in the definition of $\epsilon$ itself. Then, in the point-like limit of the detector, that is, when taking the $\epsilon \rightarrow 0$, one will arrive at an expression which is independent of the angular direction. Thus to have a direction dependence in the transition rate, one needs to have the spatial extension of the detector modelled using a finite positive $\epsilon$ parameter in the present model.

We have thus finished our construction of the direction dependent spatially extended detector. Substituting Eq.(\ref{Wfinal}) in Eq.(\ref{angtransitionrate}) gives us the angular transition rate of the detector. The expression is general and will hold for any accelerating trajectory in a flat spacetime.

\subsection{Rindler trajectory}

We shall now analyse the direction dependent transition rate for the special case of the  Rindler trajectory. 

Substituting for the trajectory $t(\tau) = (1/g)\sinh (g\tau) $, $x(\tau) =(1/g) \cosh (g\tau) $ and $y = z = 0$ in Eqs.(\ref{Tdef}) and (\ref{Xdef}), we have
\begin{align}
-T^2+X^2
& = \frac{2}{g^2} \bigg\{ 1 - \sqrt{1-{c^{\prime}}^2} \sqrt{1-{c^{\prime\prime}}^2} \cosh \bigl[ g(\tau-\tau^{\prime})-i(\alpha_{c^{\prime\prime}}+\alpha_{c^{\prime}} ) \bigr] 
\notag
\\
& \hspace{8ex}
- \left( \frac{{c^\prime}^2 + {c^{\prime\prime}}^2 }{2} \right) 
\bigg\}
\ , 
\label{TXRindler}
\end{align}
where $c^{\prime\prime}=i|\cos\Theta_{0}| g \epsilon^{\prime\prime}$, $c^{\prime}=i|\cos\Theta_{0}| g \epsilon^{\prime}$, $\cos \alpha_{c^{\prime\prime}} = 1/\sqrt{1-{c^{\prime\prime}}^2}$ and $\cos \alpha_{c^{\prime}} = 1/\sqrt{1-{c^{\prime}}^2}$. As is expected for a stationary trajectory, the above expression depends on the proper time $\tau$ and $\tau^\prime$ through their difference $\tau - \tau^\prime$ only. Further substituting Eq.(\ref{TXRindler}) in Eqs.(\ref{wfinalcompact}) and (\ref{angtransitionrate}), we get
\begin{equation}
{\dot {\cal F}}_{\Theta_0}(\omega)  =\frac{1}{16} \frac{\partial}{\partial \epsilon^{\prime}} \frac{\partial}{\partial \epsilon^{\prime\prime}} \bigg\{ 2 \Realpart \int_0^{\infty} ds \, e^{-i\omega s} \frac{4\pi\epsilon^{\prime\prime} \epsilon^{\prime}}{-T^{2}+X^{2}}  \bigg\}_{\epsilon^\prime = \epsilon, \epsilon^{\prime \prime} = \epsilon} 
\label{angFinter}
\end{equation}
Identifying the symmetry in the integrand under the simultaneous exchange of $s \rightarrow -s$ and $i \rightarrow -i$ using Eq.(\ref{TXRindler}), we can express the integral as a contour integral over the full real line $s \rightarrow (-\infty, \infty)$. Further, $(- T^2 + X^2)$ has the periodicity in $s \rightarrow s + 2\pi i /g$. One can then close the contour at $s + 2\pi i /g$ and evaluate the residue at the poles, to get 
\begin{align}
{\dot {\cal F}}_{\Theta_0}(\omega)
&= 
\frac{\partial}{\partial \epsilon^{\prime}} \frac{\partial}{\partial \epsilon^{\prime\prime}} \bigg\{\frac{\pi^{2}g\epsilon^{\prime} \epsilon^{\prime\prime}e^{\frac{\omega}{g} \left[ \tan^{-1} \left( |\cos\Theta_{0}|g\epsilon^{\prime} \right)+ \tan^{-1}\left( |\cos\Theta_{0}|g\epsilon^{\prime \prime} \right)\right]} }{4 \left( e^{\frac{2 \pi \omega}{g}}-1\right)} 
\notag 
\\[1ex]
& \hspace{13ex}
\times \frac{\sin \left( \frac{\omega}{g} \cosh^{-1} (c) \right)}{\sqrt{1-{c^{\prime}}^2} \sqrt{1-{c^{\prime \prime}}^2} \sinh \left[ \cosh^{-1} (c) \right] }   \; \;   \bigg\}_{\epsilon^\prime = \epsilon, \epsilon^{\prime \prime} = \epsilon}
\end{align}
where $c = [1 - ({c^\prime}^2 + {c^{\prime \prime}}^2)/2]/\sqrt{1 - {c^\prime}^2}\sqrt{1- {c^{\prime \prime}}^2}$. Note that $c\ge1$. 
Considering only the factor inside the braces (without the partial derivatives), it does seem to satisfy the KMS condition with the inverse of the temperature being $2 \pi/ g - (2/g)\tan^{-1} \left( |\cos\Theta_{0}|g\epsilon^{\prime} \right)+ (2/g)\tan^{-1}\left( |\cos\Theta_{0}|g\epsilon^{\prime \prime} \right)$. One gets such a result in the Schlicht case for the total transition rate with $\epsilon^{\prime} = \epsilon^{\prime \prime}$ and without the $\cos\Theta_0$ dependence. However, in the present case, the additional partial derivatives break the KMS property for ${\dot {\cal F}}_{\Theta_0}(\omega) $. 

Another way to approach at the final expression is to first differentiate the integrand in Eq.(\ref{angFinter}) and then perform the contour integration. This leads to the following
\begin{equation}
{\dot {\cal F}}_{\Theta_0}(\omega) = 2 \Realpart \int_0^{\infty} ds \, e^{-i\omega s} \, \frac{1}{D_\epsilon(s)}
\end{equation}
where 
\begin{align}
\frac{1}{D_\epsilon(s)} = \frac{ g^{2} \pi\bigg\{  3 b^2 \epsilon^2 + b^4 \epsilon^4 - 2(1- b^2 \epsilon^2)\sinh^{2}\left[ \frac{gs}{2}-i\alpha \right] - 2 i b \epsilon \sinh\left[ gs - i2\alpha \right]  \bigg\}}{32 \left( 1+b^2\epsilon^2 \right)^3 \sinh^{4}\left[ \frac{gs}{2}-i\alpha \right]}
\label{denominator}
\end{align}
and $b = g |\cos\Theta_0|$. This contour integral can be calculated using the similar procedure outlined for integral in Eq.(\ref{angFinter}) for each of the three terms. One finally gets the angular transition rate to be 
\begin{eqnarray}
{\dot {\cal F}}_{\Theta_0}(\omega) &=&  \frac{\pi g^{2}}{32 {\left(1+b^2 \epsilon^2 \right)}^3}  \; \;  \frac { e^{\frac{2\omega }{g} \tan^{-1}\left( g|\cos\Theta_{0}|\epsilon \right)}}{\left( e^{\frac{2 \pi \omega}{g}} -1\right) } \nonumber \\ 
&& \times  \bigg\{ \; \; \frac{16 \pi}{3} \left(3 b^2 \epsilon^2 + b^4 \epsilon^4 \right) \frac{\omega}{g} \left(4+\frac{\omega^2}{g^2} \right)  \nonumber \\
&& \; \; \; \; \; \; + 16 \pi \left(1-b^2\epsilon^2 \right)  \frac{\omega}{g}  + 32 \pi b \epsilon  \frac{\omega^2}{g^2}  \; \;  \bigg\}
\end{eqnarray}
Thus ${\dot {\cal F}}_{\Theta_0}(\omega)$ is not KMS thermal, in general, except when $\Theta_0 = \pi/2$. In the case $\Theta_0 = \pi/2$, $b$ vanishes and one recovers the usual Unruh temperature. Interestingly, even though the regularization in Eq.(\ref{Wfinal}) does not hold in the $\Theta_0 = \pi/2$ case, we find that the final expression is indeed finite for the case.

For $\Theta_0 \neq \pi/2$, the quadratic term in the polynomial of $(\omega/g)$ breaks the thermality of the whole expression by just a sign. One can check that the polynomial in the braces does not possess a real root and hence is positive for all real values of $\omega$. Thus the transition rate ${\dot {\cal F}}_{\Theta_0}(\omega)$ is always positive as expected. 

In the low frequency regime $|\omega/g| \ll 1$, the terms linear in $(\omega/g)$ dominate compared to the quadratic and cubic terms. Whereas, in the high frequency regime $|\omega/g| \gg 1$, the term cubic in $(\omega/g)$ dominate compared to the linear and quadratic terms. 
Hence, in both these limits, ${\dot {\cal F}}_{\Theta_0}(\omega)$ is KMS with the inverse of the temperature being equal to $2 \pi/ g - (4/g)\tan^{-1} \left( |\cos\Theta_{0}|g\epsilon \right)$, that is one observes a angle dependent temperature. In the $\Theta_0 = \pi/2$ direction, the temperature is same as the usual Unruh temperature while it increases as $\Theta_0$ decreases in the domain $0 \leq \Theta_0 \leq \pi/2$. Along the direction of acceleration, it is the maximum. 

Let us look at the combination $\epsilon |\cos\Theta_0|$. As mentioned earlier, $\epsilon$ and $|\cos\Theta_0|$ always appear as a product in the expression. Let us assume that for $\Theta_0 \neq \pi/2$, the product $\epsilon |\cos\Theta_0| \gg 1$ by assuming $\epsilon \gg 1$ for a finite $|\cos\theta_0|$. In this case, the term with pre-factor $b^4 \epsilon^4$ dominates over rest of the terms in the braces and ${\dot {\cal F}}_{\Theta_0}(\omega)$ is KMS thermal with the inverse of the temperature being equal to $2 \pi/ g - (4/g)\tan^{-1} \left( |\cos\Theta_{0}|g\epsilon \right)$. 
The $\epsilon$ parameter represents the length scale of the spatial extension of the detector. A large $\epsilon$ would signify a sufficiently extended detector. Whereas, as mentioned earlier, in the point-like limit of the detector $\epsilon \rightarrow 0$, one recovers the usual isotropic Unruh temperature. This suggests that the features mentioned above are especially due to the spatial extension of the detector.

We comment on this spurious result in the discussion section after analysing the spatially extended detector from the Rindler co-moving frame of reference in the next section. 

One might suspect that for the direction dependence feature of the spatially extended detector to vanish in the point-like limit $\epsilon \rightarrow 0^+$, could possibly be a feature of the transition rate of the detector wherein one has formally subtracted an infinite constant term by taking the difference of the transition probability of the detector at $\tau$ and $\tau + d\tau$ to arrive at the transition rate expression. The constant infinite term may, perhaps, contain the information about the direction dependence $\Theta_0$ even in the point-like limit $\epsilon \rightarrow 0^+$.

One can check for the above suspicion by explicitly computing the transition probability for the extended detector when the detector is switched ON and switch OFF smoothly \textit{around} the finite proper time $\tau_0$ and $\tau_f$ respectively. The general expression for the transition probability is given as 
\begin{equation}
{\cal F}(\omega) = \int_{-\infty}^{\infty} du \, \chi(u)  \int_{-\infty}^{\infty} ds \, \chi(u -s) e^{- i \omega s} \, W_{\Theta_0, \epsilon}(u,u-s)
\label{transitionprobability}
\end{equation}
where $\chi(\tau)$ is the smooth switching function which vanishes for $\tau < \tau_0$ and $\tau > \tau_f$ while it is unity for $\tau_0 < \tau < \tau_f$ and is smooth in its transition. For the stationary Rindler trajectory, we have $W_{\Theta_0, \epsilon}(u,u-s) = W_{\Theta_0, \epsilon}(s)$ as shown in Eqs.(\ref{TXRindler}) and (\ref{wfinalcompact}). Hence in the above equation Eq.(\ref{transitionprobability}) for transition probability, we can interchange the sequence of integration and perform the $u$ integral first to get 
\begin{equation}
{\cal F}(\omega) =  \int_{-\infty}^{\infty} ds \, e^{- i \omega s} \, W_{\Theta_0, \epsilon}(s) \, Q(s)
\label{transitionprobability2}
\end{equation}
where $Q(s) = \int_{-\infty}^{\infty} du \, \chi(u) \chi(u -s)$ is also a smooth analytic function in complex $s$ plane. One can then perform the contour integral in Eq.(\ref{transitionprobability2}), by choosing an appropriate contour and evaluating the residues of the expression at the poles of the Wightman function $W_{\Theta_0, \epsilon}(s)$ to obtain a finite result. However, from the expression in Eq.(\ref{denominator}), one can see that the combination of $\epsilon$ and $|\cos{\Theta_0}|$ always appears as product, hence evaluating the contour integral in Eq.(\ref{transitionprobability2}) would preserve the product structure which would imply that taking the point-like limit $\epsilon \rightarrow 0^+$ would make the $\Theta_0$ dependence to go away. In-fact, one can also verify that even in the non-stationary case, the product feature of the $\epsilon$ and $|\cos{\Theta_0}|$ would still hold and the direction dependence would vanish again in the the point-like limit $\epsilon \rightarrow 0^+$.

\section{A spatially extended detector in the Rindler frame}\label{rindlerframesection}

Our aim in this section is to investigate the response of an spatially extended detector, working in its co-moving frame and coupled to the Minkowski vacuum state of the scalar field, following De Bievre and Merkli~\cite{DeBievre:2006pys}. We consider the corresponding centre of mass, with co-ordinates $(x_0(\tau))$, of the detector to follow the Rindler trajectory with uniform acceleration $g$. We work in Rindler co-ordinates with the following form of the metric
\begin{equation}
ds^2 = \exp{(2 g z)} \left( - dt^2 + dz^2 \right) + d x^2_{\perp}
\label{rindlermetric}
\end{equation}

We further assume the usual monopole interaction Hamiltonian term proportional to the value of the field on the trajectory, but now the field is replaced by the smeared field $\phi(\tau)$ obtained through 
\begin{equation}
\phi(\tau) = \int dz d^2 x_{\perp} e^{g z} f \left(z, x_{\perp}, z_0(\tau), x_{\perp 0}(\tau) \right) \phi(x)
\label{smeared}
\end{equation}
where $dz d^2 x_{\perp} e^{g z}$ is the 3-spatial volume of the $t = $ constant hypersurface and  $f \left(z, x_{\perp}, z_0(\tau), x_{\perp 0}(\tau) \right)$ is the profile function which encodes the spatial geometry of the extended detector itself. For the particular detector considered, $z_0(\tau) =0 = x_{\perp 0}(\tau) $, the detector is centred, with its centre of mass at the origin.

The pullback of the Wightman function relevant for calculating the detector response function is 
\begin{eqnarray}
W(\tau,\tau^\prime) = \langle 0_M | \phi(\tau) \phi(\tau^\prime) | 0_M \rangle
\label{whitmannfunction}
\end{eqnarray}
with the transition rate being 
\begin{equation}
{\dot {\cal F}}(E) =  2 \Realpart \int_0^\infty ds \,e^{-i E s} \, W(\tau,\tau - s)
\end{equation}

The quantised scalar field in terms of the mode solutions for the metric in Eq.(\ref{rindlermetric}) is 
\begin{equation}
\phi(x) = \int d\omega \int d^2 k_{\perp} \left[ {\hat a}_{\omega, k_{\perp}}  v_{\omega, k_{\perp}}(x) + {\hat a}^{\dagger}_{\omega, k_{\perp}}  v^{\star}_{\omega, k_{\perp}}(x) \right] 
\label{field}
\end{equation}
where the mode solutions are given in terms of the modified Bessel function as
\begin{equation}
v_{\omega, k_{\perp}}(x) = \sinh \left[\frac{\pi\omega /g}{4\pi^{4}g} \right]^{1/2} K_{i\omega /g} \left[ \frac{\sqrt{k_{\bot}^{2} + m^{2}}}{g e^{-g z}} \right] e^{ik_{\perp} \cdot x_{\perp} - i \omega t} 
\label{modsol}
\end{equation}
The smeared field operator defined in Eq.(\ref{smeared}) can then be expressed as  
\begin{equation}
\phi(\tau) = \int d\omega \int d^2 k_{\perp} \left[ {\hat a}_{\omega, k_{\perp}}  h_{\omega, k_{\perp}}(\tau) + {\hat a}^{\dagger}_{\omega, k_{\perp}}  h^{\star}_{\omega, k_{\perp}}(\tau) \right] 
\end{equation}
with the corresponding smeared field modes to be 
\begin{eqnarray}
h_{\omega, k_{\perp}}(\tau) &=& \sinh \left[\frac{\pi\omega /g}{4\pi^{4}g} \right]^{1/2}  e^{ - i \omega t} \;  u_{\omega, k_{\perp}}\left(z_0(\tau), x_{\perp 0}(\tau) \right) 
\end{eqnarray}
and
\begin{eqnarray}
u_{\omega, k_{\perp}}\left(z_0(\tau), x_{\perp 0}(\tau) \right)  &=& \int dz \, d^2x_{\perp} e^{g z} f \left(z, x_{\perp}, z_0(\tau), x_{\perp 0}(\tau) \right) \nonumber \\
&& \times  K_{i\omega /g} \left[ \frac{\sqrt{k_{\bot}^{2} + m^{2}}}{g e^{-g z}} \right] e^{ik_{\perp} \cdot x_{\perp}} 
\end{eqnarray}
The pullback of the Wightman function given in Eq.(\ref{whitmannfunction}) is then expressed in terms of the smeared field modes to become
\begin{eqnarray}
W(\tau,\tau^\prime) &=& \int d\omega \int d^{2}k_{\perp} \left[ \left( \eta_{\omega}+1 \right)  h_{\omega, k_{\perp}}(\tau) h^{\star}_{\omega, k_{\perp}}(\tau^\prime) + \eta_{\omega} \,  h^{\star}_{\omega, k_{\perp}}(\tau) h_{\omega, k_{\perp}}(\tau^\prime)  \right] \nonumber \\
&=& \int d\omega \int d^{2}k_{\perp} \sinh \left[\frac{\pi\omega /g}{4\pi^{4}g} \right]   \bigg[ \left( \eta_{\omega}+1 \right)  u_{\omega, k_{\perp}}(\tau) u^{\star}_{\omega, k_{\perp}}(\tau^\prime) e^{- i \omega (\tau - \tau^\prime)} \nonumber \\
\; \; \; & & + \; \eta_{\omega} \,  u^{\star}_{\omega, k_{\perp}}(\tau) u_{\omega, k_{\perp}}(\tau^\prime) \, e^{ i \omega (\tau - \tau^\prime)}  \bigg] 
\end{eqnarray}
where $\eta_{\omega}  = 1/(\exp{(\beta \omega)} - 1)$ being the Planckian factor with the usual Unruh temperature. Since for the Rindler trajectory, we have $z_0(\tau) =0 = x_{\perp 0}(\tau) $, hence the $u_{\omega, k_{\perp}}$ are just constants. Then $W(\tau, \tau^\prime) = W(\tau - \tau^\prime) = W(s)$ as expected for a Killing trajectory. The transition rate is straightforward to obtain and we get, 
\begin{eqnarray}
{\dot {\cal F}}(E) 
&=& \int d\omega \int d^{2}k_{\perp} \sinh \left[\frac{\pi\omega /g}{4\pi^{4}g} \right]   \bigg[ \Theta(\omega + E) \left( \eta_{\omega}+1 \right)  u_{\omega, k_{\perp}} u^{\star}_{\omega, k_{\perp}}  \nonumber \\
\; \; \; & & + \; \Theta(\omega - E)\; \eta_{\omega} \,  u^{\star}_{\omega, k_{\perp}} u_{\omega, k_{\perp}}   \bigg] 
\end{eqnarray}
Thus, ${\dot {\cal F}}(E) $, satisfies the KMS condition for an arbitrary profile function,
\begin{equation}
\frac{{\dot {\cal F}}(E)}{{\dot {\cal F}}(-E)} = \frac{\eta_{E}}{\eta_{E}+1} = e^{- \beta E}
\end{equation}
with the usual Unruh temperature.

The above result regarding the thermality is quite general and holds for any arbitrary smooth profile function which falls off to Rindler's spatial infinity. One could even have included a direction dependent angle as in the case of Eq.(\ref{angprof}). However, the result would still be the same, since the spatial part does not contribute to the $\tau$  integral in the co-moving frame. 

\section{Discussion}\label{discsection}

We have analysed two models for a spatially extended detectors having direction dependence on the Rindler trajectory. The first model is based on Schlicht type construction with a direction-sensitive Lorentz-function profile for the smeared field operator with a characteristic length $\epsilon$ and defined in the Fermi co-ordinates attached to the uniformly accelerated trajectory. Whereas the second model has a very general direction sensitive profile for the smeared field operator but defined in the Rindler wedge corresponding to the trajectory. The transition rate for the two models were found to differ significantly when evaluated on the Rindler trajectory. In the first model, the spectrum was obtained to be anisotropic and non-KMS in general. Only in the two limits for the frequency $\omega/g \ll 1$ and $\omega/g \gg 1$ was the spectrum KMS thermal with a direction weighted temperature $2 \pi/ g - (4/g)\tan^{-1} \left( |\cos\Theta_{0}|g\epsilon \right)$. Further, for an arbitrary frequency but for $\epsilon /g \gg 1$ the spectrum if KMS thermal again with the same direction dependent temperature. In contrast, for the second model, the transition rate was found to be KMS thermal and isotropic for an arbitrary direction sensitive profile for the detector.

The reason for the discrepancy in results of the two models can be understood by analysing the tails of the profiles chosen relative to the Rindler horizon. In the first model with the Schlicht type profile, the constant time slices chosen in the Fermi co-ordinates extend all the way throughout the Rindler horizon since these slices form a subset of the Cauchy surfaces foliating the global flat spacetime. Hence the Lorentz-function profile defined on such a Cauchy surface has a tail extending much beyond the Rindler horizon which implies that the spatially extended detector is made up of constituents which leak outside the Rindler wedge. In such a case, when the proper time increases, the points at constant spatial coordinates on the orthogonal spatial hypersurfaces in the other Rindler wedge move to the past, and this casts doubt on the transition probability formula that involves the response function, since the formula is derived from time-dependent perturbation theory and involves time-ordered evolution. It is then a pure coincidence that Schlicht's derivation of the spectrum in Eq.(\ref{schlichttrans}) is KMS thermal with a higher temperature than the usual Unruh temperature, since the validity of the quantum description of the formula itself is suspect other than in the case when $\epsilon$ is set to zero wherein the detector model is restricted to the Rindler wedge with the usual Unruh temperature. However one can expect the transition rate expression derived in Eq.(\ref{angtransitionrate}) and (\ref{Wfinal}) is to be valid for detector trajectories not involving casual horizons.

On the other hand, If the support of the detector's profile is contained in the Rindler wedge, then the corresponding  transition rate is KMS in the usual Unruh temperature of the Rindler trajectory as is evident from the results of the second model of the detector defined in the Rindler wedge. This is regardless whether the peak of the spatial profile chosen coincides with the reference trajectory of the detector. The underlying reason is that the monopole interaction defines the energy gap of the detector with respect to the proper time of the reference trajectory, even when the proper time at the peak of the profile may be quite different from the proper time at the reference trajectory, which is the Rindler trajectory in the present case. One could question whether this way to define the interaction is a
reasonable model of the microphysics of an extended body. When the profile has a peak, perhaps a more reasonable model would be to
choose the reference trajectory to coincide with the peak of the profile. Thus, based on the second model defined in the Rindler wedge, we conclude that the Unruh effect is directionally isotropic with the usual Unruh temperature for spatially extended direction dependent detectors.

Nevertheless, there could be some interest in getting a more quantitative control of what happens for the Schlicht type detector model when the profile leaks outside the Rindler wedge, beyond Schlicht's Lorentz-function profile. Schlicht's profile gives KMS spectrum for the case discussed in the section \ref{schlichtsection} but at a different temperature; greater than the usual Unruh temperature. One could question whether other leaking profiles still give KMS (at some temperature) or does any deviation from Schlicht's profile necessarily break KMS.

\section*{Acknowledgments}

We thank Jorma Louko for helpful discussions and useful comments on the draft. 
SK thanks the Department of Science and Technology, India for
financial support and the University of Nottingham 
for hospitality where part of the work was completed.

\end{document}